\documentclass[preprint,manuscript]{aastex}
\usepackage{apjfonts,natbib}
\tighten

\def\kms{\ifmmode{~{\rm km~s^{-1}}}\else{~km s$^{-1}$}\fi}
\def\lsim{\lower0.3em\hbox{$\,\buildrel <\over\sim\,$}}
\def\gsim{\lower0.3em\hbox{$\,\buildrel >\over\sim\,$}}

\def\h2{H$_2$}
\def\flux{erg~cm$^{-2}$~s$^{-1}$}
\def\xlum{erg~s$^{-1}$}

\def\arcsec{\mbox{$^{\prime\prime}$}}

\def\umag{$U$}
\def\bmag{$B_{435}$}
\def\vmag{$V_{606}$}
\def\imag{$i_{775}$}
\def\zmag{$z_{850}$}
\def\color{$L_{60 \; \mu \mbox{\scriptsize{m}}}/L_{100 \;\mu \mbox{\scriptsize{m}}}$}
\def\udrops{$U$-dropouts}
\def\bdrops{$B_{435}$-dropouts}
\def\vdrops{$V_{606}$-dropouts}
\def\idrops{$i_{775}$-dropouts}

\def\sfr{$M_{\odot}$ yr$^{-1}$}
\def\usfr{85--240~$M_{\odot}$~yr$^{-1}$}
\def\ratio{$L_{\mbox{\scriptsize{X}}}/L_B$}
\def\Lx{$L_{\mbox{\scriptsize{X}}}$}
\def\Amin{0.1}
\def\Cmin{2.7}
\def\numu{449}
\def\numb{1734}
\def\numbrightb{395}
\def\numv{629}
\def\numi{247}


\shortauthors{LEHMER ET AL.}
\shorttitle{X-RAY PROPERTIES OF LYMAN BREAK GALAXIES}

\begin{document}

\title{X-ray Properties of Lyman Break Galaxies in the Great Observatories
Origins Deep Survey}

\author{
B.~D.~Lehmer,\altaffilmark{1}
W.~N.~Brandt,\altaffilmark{1}
D.~M.~Alexander,\altaffilmark{2}
F.~E.~Bauer,\altaffilmark{2}
C.~J.~Conselice,\altaffilmark{3}
M.~E.~Dickinson,\altaffilmark{4}
M.~Giavalisco,\altaffilmark{5}
N.~A.~Grogin,\altaffilmark{6}
A.~M.~Koekemoer,\altaffilmark{5}
K.~S.~Lee,\altaffilmark{5,6} 
L.~A.~Moustakas,\altaffilmark{5}
\& D.~P.~Schneider\altaffilmark{1}
}

\altaffiltext{1}{Department of Astronomy \& Astrophysics, 525 Davey Lab,
The Pennsylvania State University, University Park, PA 16802.}
\altaffiltext{2}{Institute of Astronomy, Madingley Road, Cambridge, CB3 0HA, United Kingdom.}
\altaffiltext{3}{Palomar Observatory, California Institute of Technology, Pasadena, CA 91125}
\altaffiltext{4}{NOAO, 950 N. Cherry Ave., Tucson AZ 85719}
\altaffiltext{5}{Space Telescope Science Institute, 3700 San Martin Drive, Baltimore, MD 21218}
\altaffiltext{6}{Johns Hopkins University, 3400 North Charles Street, Baltimore, MD 21218-2686}

\begin{abstract}

\ \ \ \ We constrain the X-ray emission properties of Lyman break galaxies
(LBGs) at $z$~$\approx$~3--6 using the $\approx$~2 Ms {\it Chandra} Deep
Field-North and $\approx$~1 Ms {\it Chandra} Deep Field-South.  Large samples
of LBGs were discovered using $HST$ as part of the Great Observatories Origins
Deep Survey (GOODS).  Deep optical and X-ray imaging over the GOODS fields have
allowed us to place the most significant constraints on the X-ray properties of
LBGs to date.  Mean X-ray properties of \numu, \numb, \numv, and \numi \ LBGs
with $z$ $\sim$~3, 4, 5, and 6, respectively, were determined using stacking
techniques. When stacked, we detect X-ray emission from LBGs at $z$~$\sim$~3
($\sim$~7$\sigma$) and from an optically bright subset (brightest 25$\%$) of
LBGs at $z$~$\sim$~4 ($\sim$~3$\sigma$); the latter is the highest redshift
detection yet for ``normal'' galaxies in the X-ray band.  The effective
exposure times for these stacked observations are $\approx$~0.7 and 0.5 Gs,
respectively.  The derived average rest-frame 2.0--8.0 keV luminosities are 1.5
and 1.4 $\times$ 10$^{41}$ \xlum, respectively.  X-ray emission from these LBGs
is likely due to high mass X-ray binaries (HMXBs) and Type II supernovae; the
corresponding star formation rates are $\approx$~\usfr.  The X-ray to $B$-band
mean luminosity ratio (\ratio) at $z$ $\sim$ 3 is somewhat elevated with
respect to that measured for starburst galaxies in the local Universe
(significance $\sim$~3$\sigma$).  When stacking full samples of LBGs at $z$
$\sim$~4, 5, and 6 we do not obtain significant detections ($<$~3$\sigma$) and
derive rest-frame 2.0--8.0~keV luminosity upper limits (3$\sigma$) of 0.9, 2.8,
and 7.1 $\times$ 10$^{41}$~\xlum, respectively.  These upper limits constrain
any widespread AGN activity in these objects to be modest at best.
Furthermore, we find that $\sim$ 0.5$\%$ of our LBGs from $z$~$\approx$~3--6
are detected individually in the X-ray band.  These LBGs have spectral shapes
and luminosities characteristic of moderate-power AGN (e.g., Seyfert galaxies
and quasars).   

\end{abstract}


\section{Introduction}\label{intro}

Determination of the basic X-ray properties of ``normal'' (i.e., not hosting a
luminous AGN) galaxies out to high redshift has been motivated in part by
\citet[hereafter GW01]{Ghosh01} who developed a model of the evolution of X-ray
luminosity with redshift.  GW01 predict that, globally, X-ray emission from
normal galaxies should peak at \hbox{$z$~$\approx$~1.5--3} due to a maximum in the
global star formation rate (SFR) at $z$~$\approx$~2.5--3.5
\citep[e.g.,][]{Blain99}.  Heightened X-ray emission at these redshifts is
expected due to an increase in the global population of low mass X-ray binaries
(LMXBs).  LMXBs evolve on timescales of $\sim$~1 Gyr, and therefore their X-ray
signatures should lag behind more immediate tracers of star formation.

Stacking analyses and deep X-ray surveys with the {\it Chandra X-ray
Observatory} (hereafter {\it Chandra}) have enabled testing of these
predictions out to cosmologically interesting distances.  Stacking techniques
generally involve the addition of X-ray counts at known positions of galaxies
to yield average X-ray detections where individual sources lie below the
detection threshold.  For example, \citet[hereafter H02]{Horn02} used stacking
with a $\approx$ 1~Ms exposure of the {\it Chandra} Deep Field-North (CDF-N) to
investigate the evolution of the X-ray luminosities of \hbox{$z$ = 0.4--1.5}
spiral galaxies.  This analysis shows suggestive evidence for evolution in
X-ray luminosity with redshift at a level somewhat lower than that predicted by
GW01.  H02 also found that the X-ray to $B$-band luminosity ratio, \ratio, is
elevated for $z$~$\approx$~1 spirals with rest-frame $L_B$~$\approx$~$L_B^*$ as
compared with galaxies in the local Universe.  \ratio \ has been shown to be
linked to star formation activity.  This is supported by the existence of an
empirical correlation between \ratio \ and \color \ \citep[][]{Fabbiano2002}.
Here, the quantity \color \ is a measure of far infrared color temperature,
which has been shown to increase with star formation activity
\citep[e.g.,][]{Helou1991}.  More recent stacking analyses of large samples of
galaxies in the $\approx$ 2~Ms CDF-N and $\approx$ 1~Ms {\it Chandra} Deep
Field-South (CDF-S) show that the X-ray properties of normal galaxies of all
morphological types evolve similarly from $z$~=~0.4--1.5 \citep{Wu2004}. 

At higher redshifts ($z$~$\approx$~2--6) stacking analyses have been performed
using individually undetected normal galaxies (e.g., Brandt et al.~2001,
hereafter B01; Nandra et al.~2002, hereafter N02; Malhotra~et~al.~2003; Bremer
et al.~2004; Moustakas \& Immler~2004 Reddy \& Steidel~2004;
Wang~et.~al.~2004).  At $z$~$\sim$~3 these galaxies often have mean X-ray
luminosities and \ratio \ comparable to bright starburst galaxies in the local
Universe.  The heightened X-ray emission from these galaxies is probably due to
newly formed high mass X-ray binaries (HMXBs) and young supernova remnants.
HMXBs have much shorter formation timescales than LMXBs and are therefore more
immediate tracers of cosmic star formation history.  

Recently, large samples of galaxies have been identified at $z$ $\sim$ 3, 4, 5,
and 6 as part of the Great Observatories Origins Deep Survey (GOODS).  These
galaxies were isolated via the Lyman break technique (e.g.,
Steidel~et~al.~1995; Madau~et~al.~1996; Steidel~et~al.~1999), which estimates
the redshift of a galaxy based on its ``dropout'' bandpass (see
$\S$~\ref{samples}).  Lyman break galaxies (LBGs) generally show rest-frame
UV/optical characteristics similar to those of local starburst galaxies---weak
or absent Lyman $\alpha$ emission, P~Cygni features from C~{\small IV} and
other lines, and strong interstellar UV absorption lines such as Si~{\small
II}, O~{\small I}, C~{\small II}, Si~{\small IV}, and Al~{\small II} (see e.g.,
Steidel et al. 1996; Shapley et al. 2003).  The GOODS covers
$\approx$~316~arcmin$^2$ of sky in two fields, \hbox{GOODS-North}
(\hbox{GOODS-N}) and \hbox{GOODS-South} (\hbox{GOODS-S}) \citep{Giav2004a}.
Both fields are subregions of the {\it Chandra} Deep Fields for which {\it
Chandra} has acquired an $\approx$ 2~Ms observation of the CDF-N
\citep[hereafter A03]{Alexander2003a} and an $\approx$ 1~Ms observation of the
{\it Chandra} Deep Field-South (CDF-S; Giacconi et al.  2002; A03).

In this paper, we use stacking techniques to constrain the X-ray properties of
LBGs in the GOODS fields.  We improve on results from B01 and N02 by
increasing the number of galaxies at $z$~$\sim$~3 from 24 (B01) and 148 (N02)
to \numu, and we add to this samples of \numb, \numv, and \numi \ galaxies at
$z$ $\sim$ 4, 5, and 6, respectively.

The Galactic column densities are 1.3 $\times$ 10$^{20}$~cm$^{-2}$
\citep{Lockman2003} and 8.0 $\times$ 10$^{19}$~cm$^{-2}$ \citep{Stark1992} for
the CDF-N and CDF-S, respectively.  $H_0$ = 70~\hbox{km s$^{-1}$ Mpc$^{-1}$},
$\Omega_m$ = 0.3, and $\Omega_{\Lambda}$ = 0.7 are adopted throughout this
paper \citep{Spergel2003}.  Coordinates are J2000.0.

\section{Analysis}\label{analysis}

\subsection{Samples}\label{samples}

The Lyman break technique has been utilized in the GOODS regions, where deep
mosaic images have been taken using the $HST$ Advanced Camera for Surveys (ACS)
with bandpasses \bmag, \vmag, \imag, and \zmag \ \citep[hereafter
G04]{Giav2004b}; the full-depth, five-epoch ACS images have been used here.
Additional ground-based photometry has been obtained over the GOODS fields in
the \umag \ band at KPNO \citep[\hbox{GOODS-N}]{Cap2004} and CTIO
(\hbox{GOODS-S}).  Identification of these LBGs is based on the following color
equations:\\ 
\udrops:
$$(U - B_{450}) \ge 0.75 + 0.5 (B_{450} - z_{850}) \; \; \wedge $$
\begin{equation}
(U - B_{450}) \ge 0.9 \; \; \wedge \; \; (B_{450} - z_{850}) \le 4.0
\end{equation}
\bdrops:
$$(B_{450} - V_{606}) \ge 1.2 + 1.4 \times (V_{606} - z_{850}) \; \; \wedge $$
\begin{equation}
(B_{450} - V_{606}) \ge 1.2 \; \; \wedge \; \; (V_{606} - z_{850}) \le 1.2
\end{equation}
\vdrops:
$$[(V_{606} - i_{775}) > 1.5 + 0.9 \times (i_{775} - z_{850})] \;\; \vee $$
$$[(V_{606} - i_{775}) > 2.0] \;\; \wedge \;\; (V_{606} - i_{775}) \ge 1.2 \;\; \wedge $$
\begin{equation}
(i_{775}-z_{850}) \le 1.3 \;\; \wedge \; \; \mbox{S/N (} B_{450} \mbox{)} < 2.0
\end{equation}
\idrops:
$$(i_{775} - z_{850}) \ge 1.3 \; \; \wedge $$
\begin{equation}
\mbox{S/N (} B_{450} \mbox{)} < 2.0 \; \; \wedge \; \; \mbox{S/N (} V_{606} \mbox{)} < 2.0
\end{equation}

\noindent Here the symbols $\wedge$ and $\vee$ correspond to the logical
operators AND and OR, respectively.  Monte Carlo simulations of the
color-selection procedure (e.g., G04) indicate that these color criteria select
galaxies with mean redshifts and 1$\sigma$ redshift ranges of $z$ =
3.01~$\pm$~0.24, 3.78~$\pm$~0.34, 4.92~$\pm$~0.33, and 5.74~$\pm$~0.36 for
\umag-, \bmag-, \vmag-, and \idrops, respectively.  Dropout lists created in
G04 were inspected and filtered to minimize the number of interlopers and
contaminants in the samples.  The interloper fraction is predicted to be
$\approx 10 \%$ for \udrops \ and \bdrops.  This fraction increases to become
significant for \vdrops \ and \idrops \ ($\approx$ 20$\%$ and 40$\%$,
respectively; see Dickinson et al.~2004; G04) 

\begin{figure}
\centerline{
\includegraphics[width=9.5cm,angle=0]{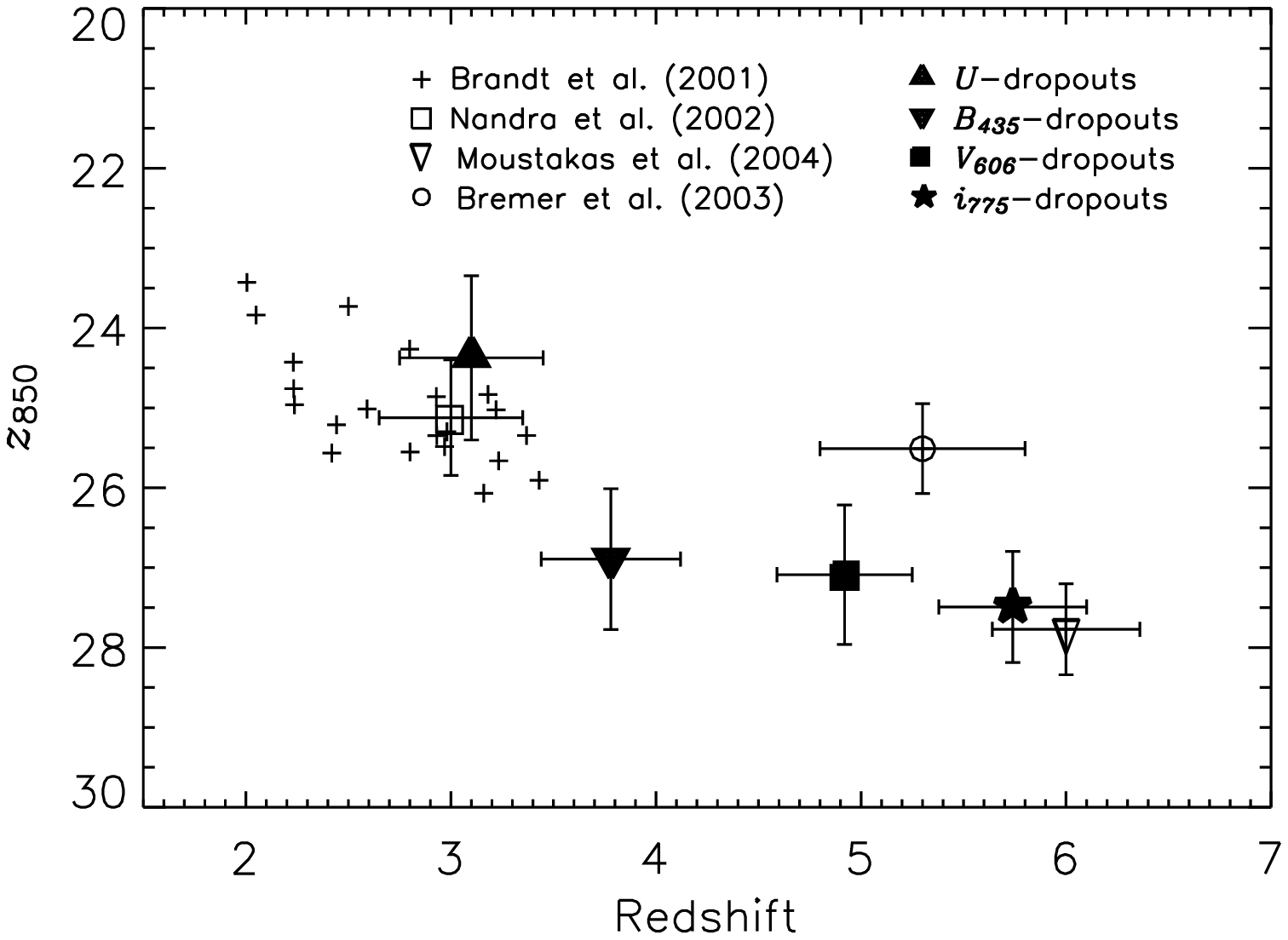}
}
\figcaption[fig1.eps]{\zmag \ magnitude vs. redshift for
X-ray stacking analyses of LBGs.  Filled (our LBGs) and open symbols (other
investigations) represent the estimated mean \zmag \ and redshift with error
bars showing the 1$\sigma$ spread for each quantity.  The Brandt et al.~(2001)
measurements (+) represent the positions of individual LBGs with
spectroscopic redshifts.  \label{fig:magvsz}}
\end{figure}

The LBGs used in our stacking analyses have average redshifts and \zmag \
magnitudes as shown in Figure~\ref{fig:magvsz}; for comparison the \zmag \
magnitudes of several other stacking analyses in this redshift range are also
plotted.  Galaxies in this survey probe relatively faint optical fluxes over a
wider range of redshifts ($z \approx 3-6$) than those used for previous X-ray
stacking analyses.  The corresponding lookback times for \umag-, \bmag-,
\vmag-, and \idrops \ are 11.5, 12.1, 12.5, and 12.7~Gyr, respectively
\citep[see ][]{Hogg2000}.

The stacking procedure used in our analyses was similar to that of B01 and
N02; the intent was to obtain average X-ray properties (e.g., luminosities and
spectral shapes).  At each LBG position, photon counts and effective exposure
times (from exposure maps, which include the effects of vignetting) were
extracted from circular apertures and summed to give a total number of
counts and a total effective exposure time.  We excluded LBGs located within
10\arcsec \ of individually detected X-ray sources in A03; this avoids
contamination by unrelated sources.  This exclusion process led to the
rejection of 99, 267, 127, and 31 \umag-, \bmag-, \vmag-, and \idrops,
respectively ($\approx$ 18$\%$, 13$\%$, 17$\%$, and 11$\%$ of the respective
LBG populations).  A small number of LBGs (18) were found to be within the
positional errors of {\it Chandra} sources; their characteristics are given
in $\S$~\ref{detected} and listed in Table~\ref{tab:detected}.  

The background was estimated through Monte Carlo analysis using a background
map (see $\S$~4.2 of A03).  Each LBG position was shifted randomly within
25$\arcsec$ of the original position, and background counts were obtained for
each new position and summed just as they were for the LBGs themselves.  This
procedure was repeated 10,000 times to obtain an accurate estimate of the
local background and its dispersion.  To verify our results were not biased by
gradients in the local background, we also tried shifting in regions with
radii of 10$\arcsec$, 15$\arcsec$, 20$\arcsec$, 30$\arcsec$, and 35$\arcsec$
and found no material differences in our results.

\begin{figure}
\centerline{
\includegraphics[width=9.5cm,angle=0]{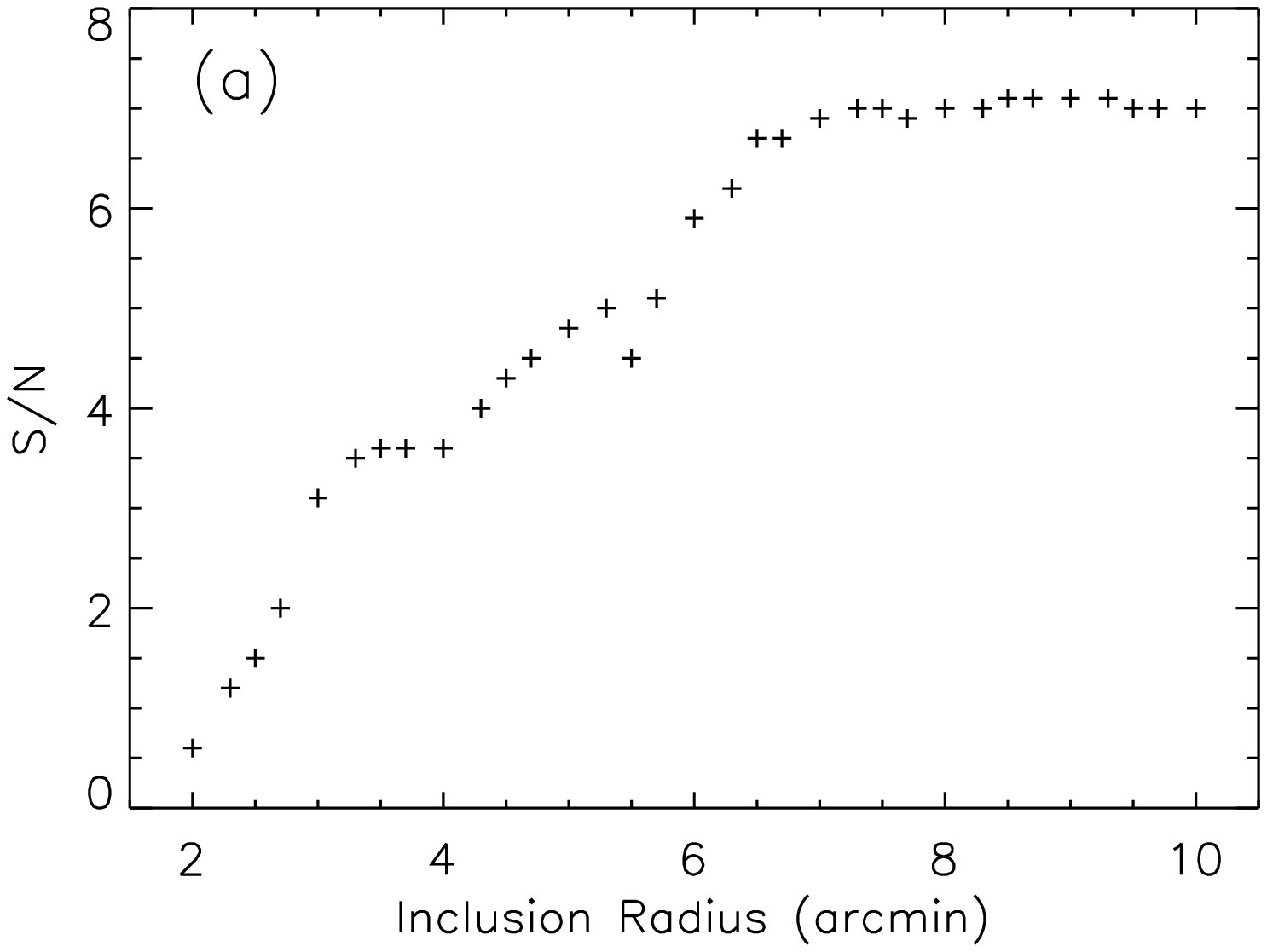}
\hfill
\includegraphics[width=9.5cm,angle=0]{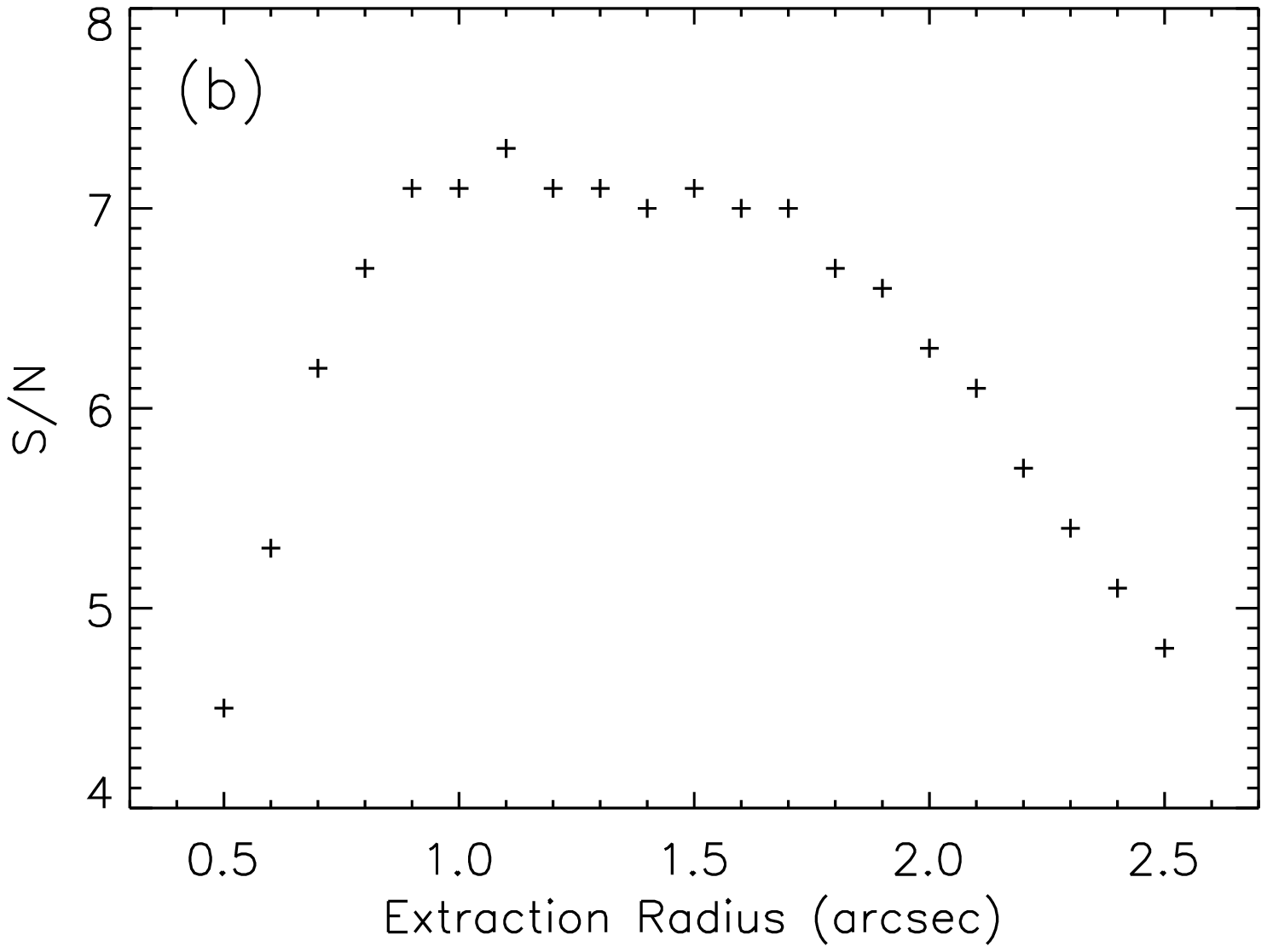}
} 
\figcaption[fig2.eps]{ ({\it a}) The achieved signal-to-noise ratio (S/N)
as a function of inclusion radius for stacking \udrops.  Note that the number
of stacked sources rises with increasing inclusion radius and is maximized at
$\approx$~9$\farcm$0.  ({\it b}) Signal-to-noise ratio (S/N) as a function of
extraction radius.  Here, the signal is strongest for an aperture radius of
\hbox{$\approx$ 1$\farcs$5}.  This process of optimizing the S/N was achieved
iteratively and appears here in convergence.  \label{fig:apr}}
\end{figure}

Our stacking procedure maximized the signal-to-noise ratio (hereafter,
S/N)\footnote{Here, S/N $\equiv$ $(S-B)/B^{0.5}$, where $S$ is the source plus
background counts and $B$ is the average background counts obtained from the
Monte Carlo procedure.  This approximation is accurate for $\vert S-B \vert \ll
B$ and $B$ $\ge$ 20, which applies to all cases in our analyses.} by varying
the aperture cell size for extracting X-ray counts and the off-axis angle
(angle between a particular source and the average CDF-N or CDF-S aim points)
within which sources were included in stacking.  This procedure was performed
using \udrops \ because of their strong X-ray signal.  We avoided stacking
sources that were outside a certain off-axis angle due to the increased size of
the point-spread function (PSF) and confusion with background in this region.
In the maximization process we chose a specific aperture cell size and stacked
sources within various off-axis angles (hereafter, inclusion radii) until S/N
was maximized.  Sources within the resulting inclusion radius were then stacked
using variable aperture cell sizes.  S/N was maximized for a specific aperture
cell size, and the process was repeated iteratively until convergence.
Figures~\ref{fig:apr}{\it a} and~\ref{fig:apr}{\it b} show the result of this
process.  We found that S/N peaks for an inclusion radius $\approx$~9$\farcm$0
and an aperture size of $\approx$~1$\farcs$5 (as expected for small off-axis
angles); these values are used throughout our analyses.

Mean count rates $\Phi$ were calculated by subtracting our best estimate of the
local background counts $B$ from the background plus source counts $S$ and
dividing by the total effective exposure time $T$ [i.e.,
\hbox{$\Phi$~=~($S-B$)/$T$}].  To convert count rate to flux we used the
Galactic column densities discussed in $\S$~\ref{intro} and adopted a power-law
photon index of $\Gamma$ = 2.0, appropriate for starburst galaxies
\citep[e.g.,][]{Kim92,Ptak1999}.  Sources were stacked in both the
\hbox{GOODS-N} and \hbox{GOODS-S} fields, so we used a statistically weighted
Galactic column density.  X-ray luminosities were calculated following
\citet{Schmidt}:

\begin{equation}\label{eqn}
L_{\mbox{\scriptsize{X}}} = 4 \pi d_L^2 \; f_{\mbox{\scriptsize{X}}} \; (1+z)^{\Gamma - 2} \; \mbox{erg s}^{-1} 
\end{equation} 

\noindent where $d_L$ is the luminosity distance (cm), and
$f_{\mbox{\scriptsize{X}}}$ is the X-ray flux (erg cm$^{-2}$ s$^{-1}$).

\subsection{Subsets}\label{subsets}

The large number of LBGs in our sample enabled investigation of the
contribution of specific subsets to the signal.  We chose to divide the
original dropout lists based on galaxy morphology (\udrops \ only) and
rest-frame $B$-band luminosity.  We used the CAS morphology system
\citep[hereafter C03]{Con2003} to divide the $U$-dropout list according to the
observed-frame \zmag-band (rest-frame $\lambda \lambda$~2000--2500)
concentration and asymmetry parameters.  We note that the CAS parameters have
been derived using rest-frame UV light and may have somewhat different physical
interpretations than those quoted for the local Universe.  Concentration ($C$)
is proportional to the logarithm of the ratio of the radii containing 80$\%$
and 20$\%$ of the source flux: 

\begin{equation} C = 5 \times
\mbox{log}\;(r_{80\%}/r_{20\%}) 
\end{equation} 

\noindent Galaxies with high $C$ values (e.g., bulge dominated systems locally)
are generally brightest in their central regions \citep{Con2002}.  Asymmetry
($A$) is defined for a galaxy by taking the absolute value of the difference in
fluxes of the galaxy's image and its 180$\degr$ rotated analog and dividing by
the original image flux (see Conselice, Bershady, \& Jangren 2000; C03, for
details).  Generally disk-dominated systems and galaxy mergers are observed to
have high $A$ values in the local Universe.  Parameter values of $A$ = \Amin \
and $C$ = \Cmin \ were chosen to divide the $U$-dropout list roughly in half,
resulting in lists of LBGs with $A$~$<$~\Amin, $A$~$>$~\Amin, $C$~$<$~\Cmin,
and $C$~$>$~\Cmin.

We also divided our original dropout lists into subsets based on rest-frame
$B$-band luminosity. Rest-frame $B$-band luminosities were calculated using a
spectral energy distribution (SED) and applying $K$-corrections to the fluxes
(e.g., Schneider, Gunn, \& Hoessel~1983; Hogg~2002).  $K$-corrected ACS
magnitudes were used in these computations, thus the derived rest-frame
$B$-band luminosities are strongly dependent on the SED choice.  An SED for
LBGs was used in calculating $K$-corrections.  This SED was derived from the
Bruzual \& Charlot (2003) solar metallicity models, with a Salpeter initial
mass function, and aged by 144 Myr at constant SFR.  Optically bright and faint
sublists were generated by dividing each dropout list (i.e., \umag-,
\hbox{\bmag-,} \vmag-, and \idrops) at its mean rest-frame $B$-band luminosity.
The corresponding mean rest-frame $B$-band luminosities used in these divisions
were $L_B$ $\approx$ 5.3, 5.5, 4.1, and 5.5~$\times$~10$^{43}$ \xlum \
($M_B$~$\approx$~$-$20.6, $-$20.5, $-$20.1, $-$20.6) for \umag-, \bmag-,
\vmag-, and \idrops, respectively.  Because rest-frame $B$-band luminosity
distributions are asymmetric with skew tails toward high luminosity, splitting
at the mean leads to two lists with unequal sizes.

\section{Results}\label{results}

\subsection{Detected Sources}\label{detected}

Over the GOODS fields seven \udrops, eight \bdrops, two \vdrops, and one
$i_{775}$-dropout \ were found to be coincident with {\it Chandra} sources
within the positional errors (see Table~\ref{tab:detected}).\footnote{Note that
the Lyman break technique is a statistical process for isolating galaxies at a
given redshift.  Due to the statistical nature of this process, we do not
expect to recover all objects within the ACS flux limits that may reside at the
redshifts under investigation here.}  Positional errors, photon counts,
effective photon indices, and fluxes for these sources were determined in A03.
The 2.0--8.0~keV X-ray luminosities were computed using
eqn.~\ref{eqn}.\footnote{For reference, the on-axis sensitivity limit for the 2
Ms CDF-N is $\approx$ 2.5 $\times$ 10$^{-17}$ \flux \ in the soft band
corresponding to rest-frame \hbox{2.0--8.0~keV} luminosities of 2.0, 4.0, 6.8,
and 10.4 $\times$ 10$^{42}$ \xlum \ at $z$ = 3, 4, 5, and 6, respectively
($\Gamma$ = 2).}  We investigated the probability of obtaining a false match by
shifting all of our LBG positions and checking to see if the new positions were
coincident with {\it Chandra} sources.  LBG positions were shifted by 10\arcsec
\ in 16 different directions, and an average of $\approx$ 2 detections were
obtained for each direction.

Of the 18 detected sources, we find that 14 have corresponding spectroscopic
\citep{Barger2003a,Cristiani2004,Szokoly04} and/or photometric
\citep{Barger2003a,Mobasher2004,Zheng2004} redshifts.  Of these 14 sources,
four (J123621.0+621412, J123642.2+620612, J123701.6+621146, J033243.2$-$274914)
have redshifts inconsistent with those determined by the Lyman break technique.
We note that the redshifts derived for three of the four sources
(J123621.0+621412, J123642.2+620612, and J123701.6+621146) are  photometric
redshifts, implying there are still uncertainties in their values.  These may
perhaps be dismissed as being ``inconsistent.''  The spectroscopic redshift of
$z$~=~1.92 for $U$-dropout J033243.2$-$274914 would therefore be the only
remaining discrepant case.  This LBG may be an interloper (see
$\S$~\ref{samples} for probabilities) or falsely matched.  Furthermore, the
colors of some AGN may satisfy the color criteria used to select normal
galaxies of differing redshifts.  This may cause a discrepancy between
spectroscopic and color-selected redshifts derived for these AGN.  

The detected sources are mostly AGN with moderate luminosities (i.e., \Lx \
$\approx$ 10$^{43}$--10$^{44.5}$ \xlum), characteristic of luminous Seyfert
type galaxies to quasars.  Several of these sources have been characterized in
previous investigations.  J123648.0+620941 is the highest redshift quasar known
($z$ = 5.186) in the CDF-N and -S fields \citep[e.g.,][]{Barger2003b,Vig2002}.
J033201.6$-$274327, J033218.2$-$275241, J033243.2$-$274914, J033244.3$-$275251
\citep{Szokoly04}, and J033209.4$-$274807 \citep{Schreier01} have been
spectroscopically classified as broad-line AGN.  J123701.6+621146 and
J123647.9+620941 are both Very Red Objects \citep[$I-K \ge$~4,
VROs;][]{Alexander2002a}.  J033229.8$-$275105 has been identified as a putative
type 2 QSO \citep{Norm2002}.  Finally, three additional sources
(J033239.7$-$274851, J033242.8$-$274702, and J033250.2$-$275252) have been
cataloged as high-redshift QSO candidates \citep{Cristiani2004}.

Following \citet{Alexander2002b} we assume detected sources with 2.0--8.0~keV
\Lx \ $>$ 1.5 $\times$ 10$^{42}$ \xlum \ or $\Gamma$ $<$ 1.0 are AGN.  Only one
source (J123642.2+620612) falls outside these criteria, leaving 17 clear AGN.
The derived AGN fractions for our \umag-, \bmag-, \vmag-, and \idrops \ are
$\approx$~1.2$\%$, 0.4$\%$, 0.3$\%$, and 0.4$\%$, respectively; these fractions
are lower than others found in previous investigations.  For comparison, the
AGN fraction derived from N02 is $\approx$~2.7$\%$.  In comparison to our
\bmag-, \vmag-, and \idrops, our \udrops \ and the LBGs used by N02 are
relatively luminous in the optical.  The corresponding AGN fractions therefore
suggest X-ray luminous AGN are generally optically luminous as well.

\subsection{Stacked Sources}\label{stacked}

\begin{figure}
\centerline{
\includegraphics[width=9.5cm,angle=0]{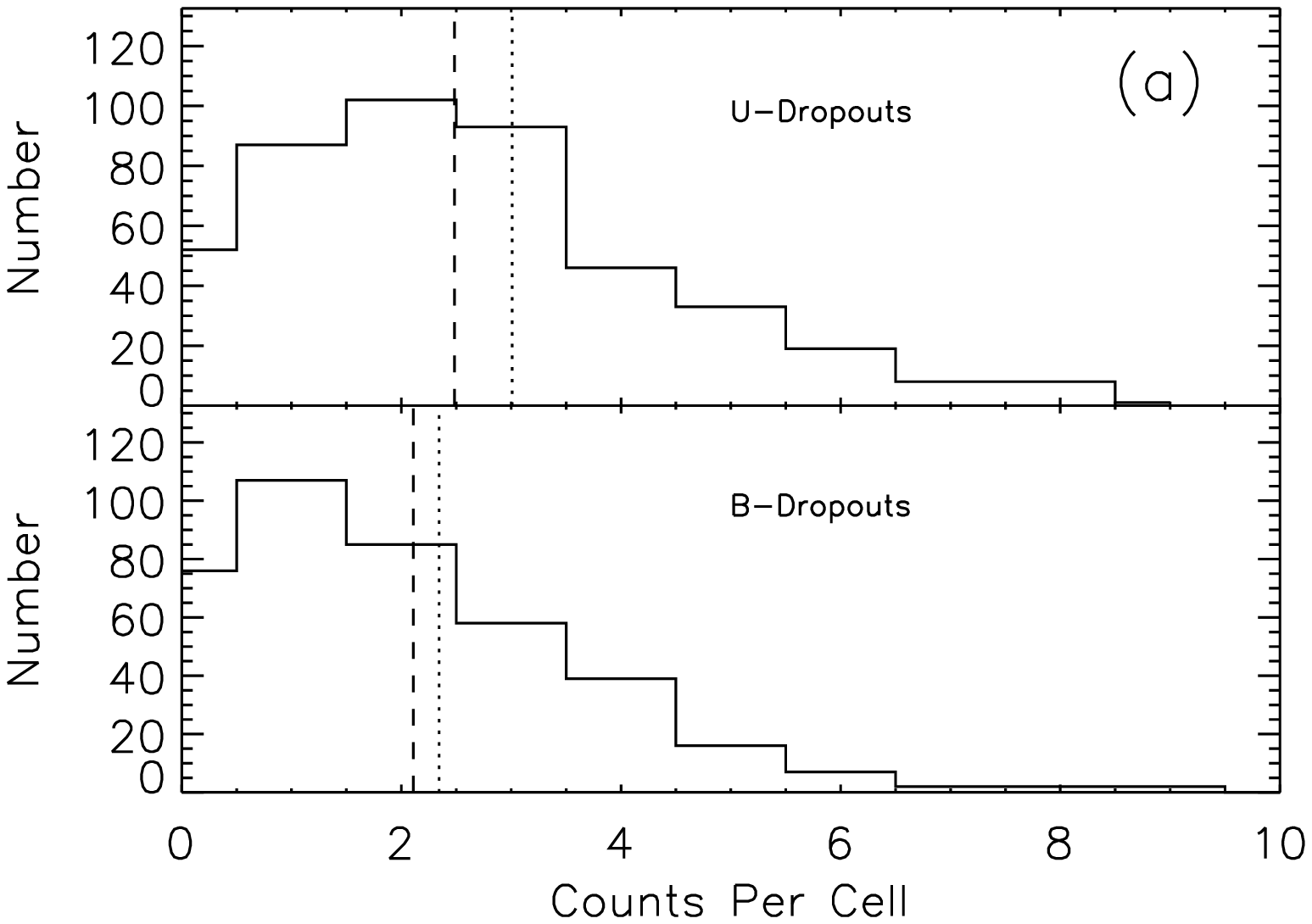}
\hfill
\includegraphics[width=9.5cm,angle=0]{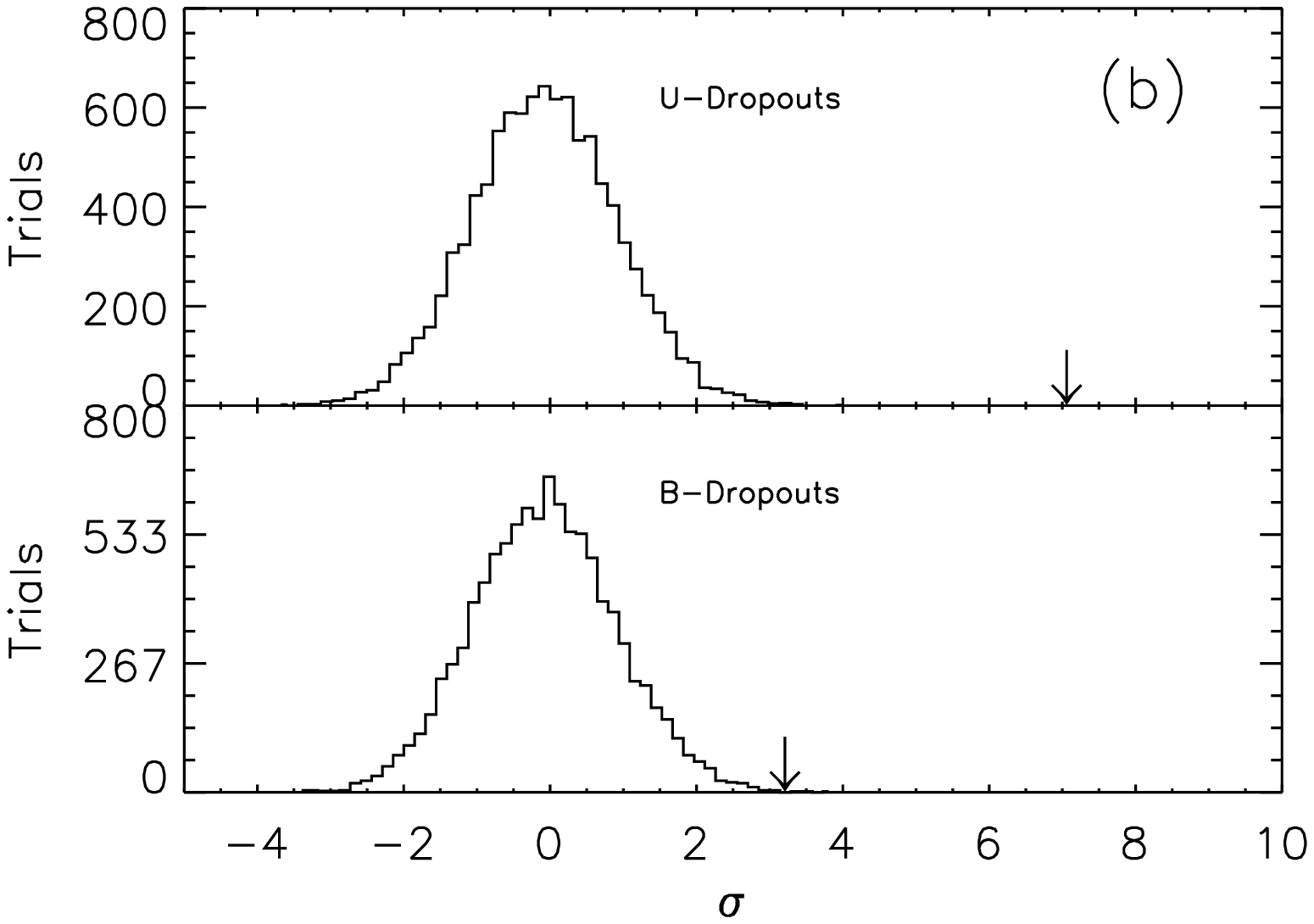}
}
\figcaption[fig3.eps]{({\it a}) Histogram of the count distribution for
\udrops \ ({\it upper panel}) and \bdrops \ ({\it lower panel}).  The vertical
dotted lines show the mean source plus background counts per aperture cell, and
the vertical dashed lines show the mean background counts per aperture cell.
({\it b}) Results from Monte Carlo estimates of the background level for
\udrops \ ({\it upper panel}) and \bdrops \ ({\it lower panel}).  The detection
level is represented with downward pointing arrows.  \label{fig:monte}}
\end{figure}

\begin{figure}
\centerline{
\includegraphics[width=9.7cm,angle=0]{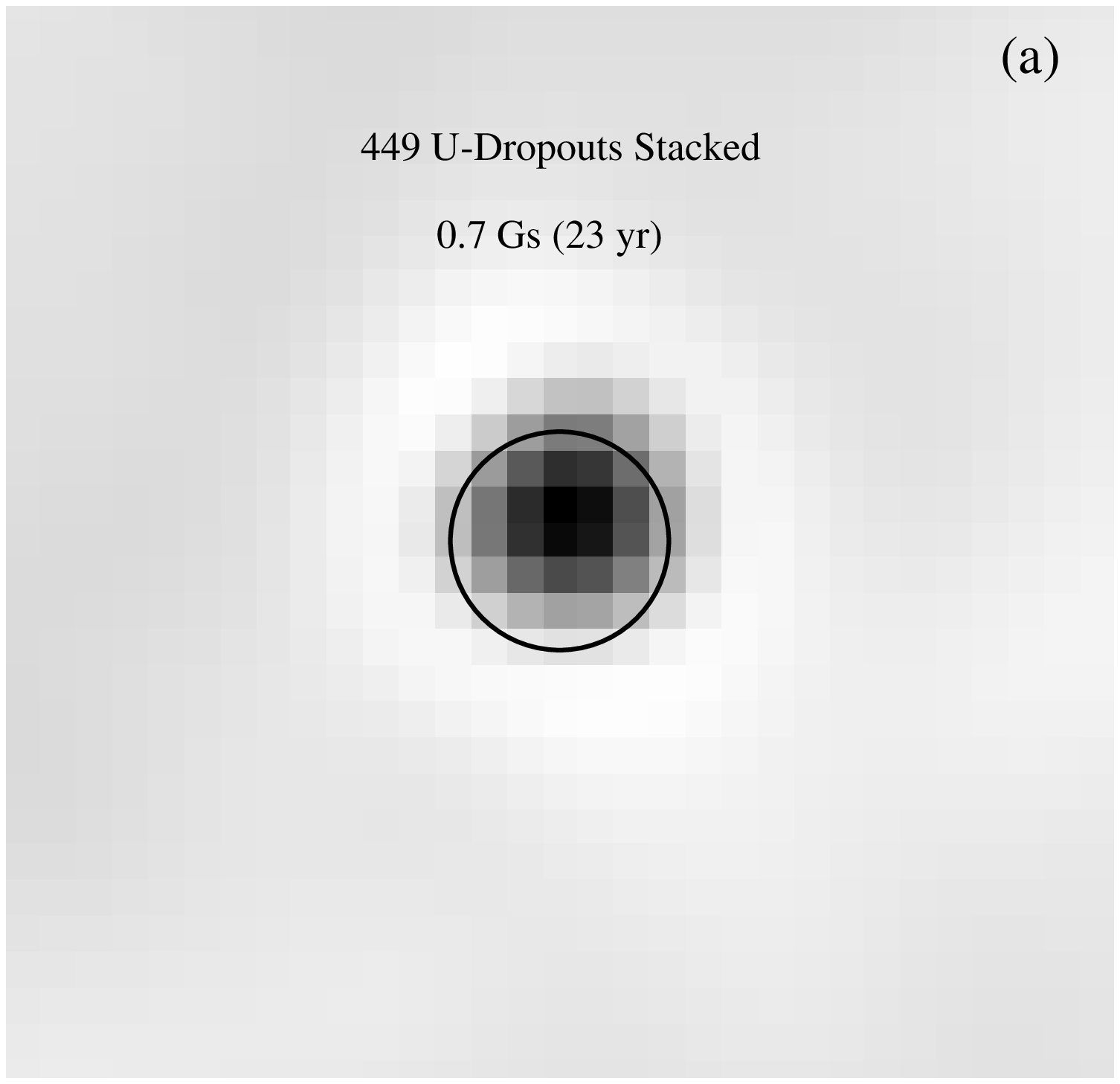}
\hfill
\includegraphics[width=9.7cm,angle=0]{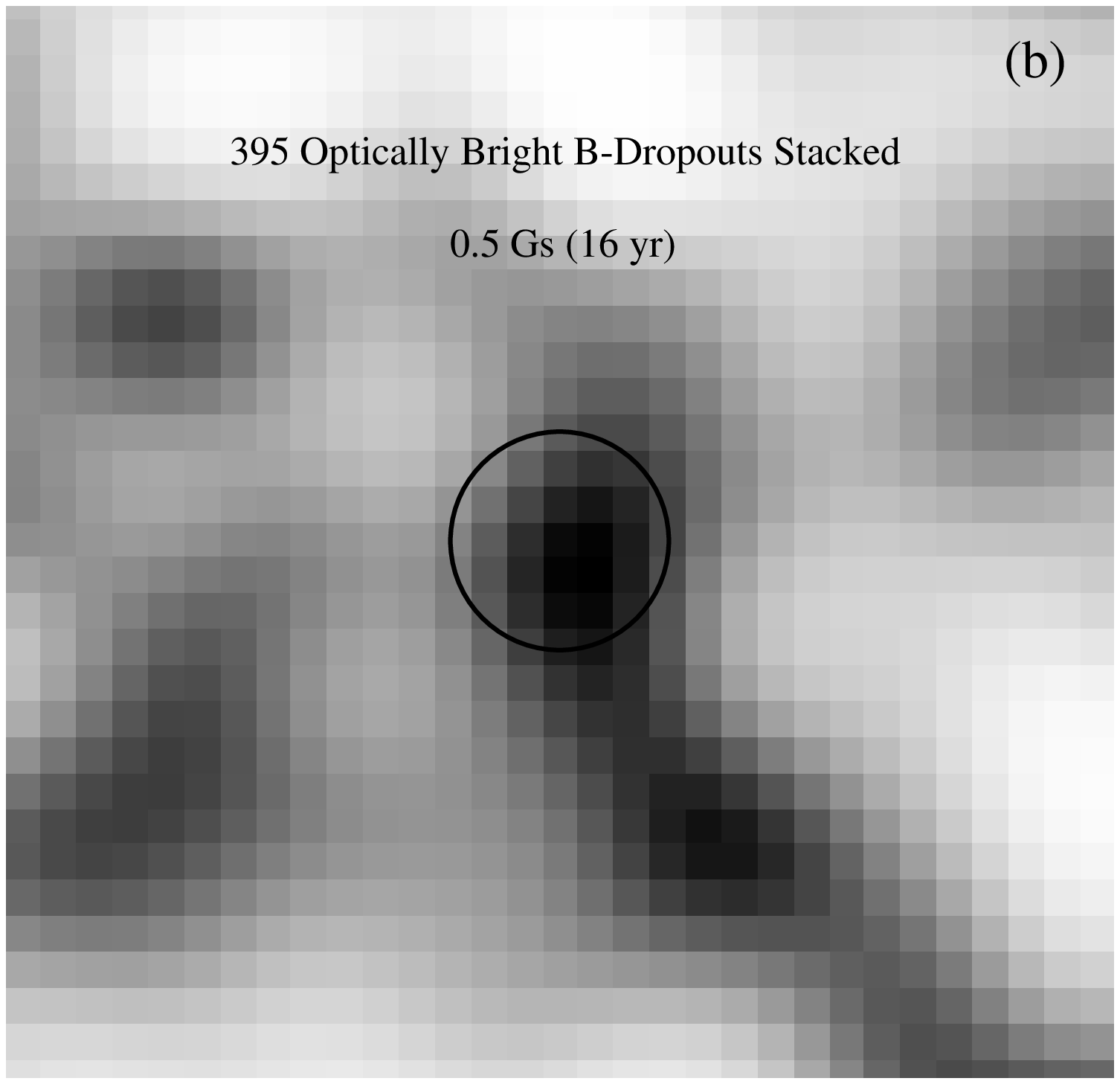}
}
\figcaption[fig4.eps]{ Stacked soft-band images of \numu \
\udrops \ ($\approx$ 0.7 Gs exposure) and \numbrightb \ optically bright
\bdrops \ ($\approx$ 0.5 Gs exposure). Stacked emission from these LBGs is
significantly detected in the soft-band with significances of $\sim$~7$\sigma$
(\udrops) and $\sim$~3$\sigma$ (bright \bdrops).  The images are 15\arcsec \
$\times$ 15\arcsec \ (0$\farcs$5 pixel$^{-1}$) and were adaptively smoothed at
2.5$\sigma$ using the CIAO tool CSMOOTH.  The faint ``nebulosity'' observed in
the optically bright \bdrops \ is attributed to smoothing over noise.  The
black circles are centered on our 1$\farcs$5 radius aperture cell that was used
in the stacking analyses.  \label{fig:images}}
\end{figure}

The results of our stacking analyses of individually undetected sources are
found in Table~\ref{tab:maglist}.  Hereafter, unless stated otherwise, we
discuss the results in reference to the soft-band stacking analyses where we
obtain significant detections; we do not obtain significant detections in the
hard- or full-bands (this result is understandable due to the significantly
lower background in the soft-band).  The stacking procedure was repeated for
the \hbox{GOODS-N} and \hbox{GOODS-S} fields individually; results from the two
fields were statistically consistent.  Generally, we detect the stacked
emission from \udrops \ ($\sim$ 7$\sigma$) and all subsets generated therefrom.
We do not obtain significant detections (S/N $>$ 3$\sigma$) for general samples
of \bmag-, \vmag-, and \idrops, but we note a suggestive positive fluctuation
($\sim$~2$\sigma$) for our \bdrops.  We do, however, obtain a significant
detection for the bright subset of \bdrops \ ($\sim$ 3$\sigma$).
Figure~\ref{fig:monte}{\it a} shows the distributions of counts obtained for
the individual galaxies being stacked for both \udrops \ and the bright subset
of \bdrops.  We obtained an average of 3.1 and 2.3 counts per cell for \umag- \
and bright \bdrops, respectively.  The summed numbers of source plus background
counts were 1351 counts for \udrops \ and 926 for bright \bdrops.  In the
10,000 Monte Carlo trials where we shifted the aperture cells to random
positions and summed background counts, we found that no trials produced
$\ga$~1351 counts for \udrops \ and only 6 trials produced $\ga$~926 counts for
\bdrops.   Gaussian statistics predict $\approx$~0 and 5 Monte Carlo trials
should exceed the total source plus background counts for our analysis of
\udrops \ and bright \bdrops, respectively.  Figure~\ref{fig:monte}{\it b}
shows the Monte Carlo distributions obtained when stacking random positions
over 10,000 trials.  In both cases the detection confidence levels are greater
than 99.9$\%$.  Stacked and smoothed images are displayed in
Figure~\ref{fig:images}.  The images have effective exposure times of $\approx$
0.7 and 0.5 Gs (22 and 16 yr) for \umag- and bright \bdrops, respectively.

Stacked \vmag- and \idrops \ and subsets thereof produced no significant
detections (i.e, S/N~$<$~3$\sigma$).  We also attempted to merge the bright
subsets of \vdrops \ and \idrops \ and failed to obtain detections.  X-ray
emission constraints for these LBGs are tabulated in Table~\ref{tab:maglist}.

\section{Discussion}\label{discuss}

Using the $\approx$ 2 Ms CDF-N plus $\approx$ 1 Ms CDF-S we have placed
constraints on the X-ray properties of LBGs identified in the 316 arcmin$^2$
\hbox{GOODS-N} and -S fields.  We have used X-ray stacking techniques on samples
of \numu, \numb, \numv, and \numi \ individually undetected LBGs at $z$ $\sim$
3, 4, 5, and 6, respectively (\umag-, \bmag-, \vmag-, and \idrops,
respectively) to obtain their average X-ray properties.  X-ray emission from
LBGs is expected to be largely due to activity associated with star-forming
processes (e.g., HMXBs and young supernova remnants) with a low-level
contribution from low-luminosity AGN \citep[LLAGN; e.g.,][]{Ho2001}.  With the
intent to investigate normal/starburst galaxies we have rejected LBGs
coincident with known {\it Chandra} sources from our stacking analyses.
Considering {\it Chandra's} superb sensitivity for detecting luminous AGN, we
are confident that this rejection process has effectively removed strong
accretion-dominated X-ray sources that would heavily contaminate our stacked
signal.  We have identified 18 LBGs coincident with {\it Chandra} sources (see
$\S$~\ref{detected}).  With the possible exception of one source we find these
objects are moderately luminous AGN comprising $\approx$ 0.5$\%$ of the LBG
population from $z \sim 3-6$.

Using our stacking procedure (see $\S$~\ref{analysis}) we detect X-ray emission
from LBGs at $z$ $\sim$ 3 and an optically bright subset of LBGs at $z$ $\sim$
4.  X-ray count rates derived from our soft-band detections and hard-band upper
limits (3$\sigma$) were used to constrain the spectral slopes of these LBGs.
The \hbox{2.0--8.0 keV} to \hbox{0.5--2.0 keV} band ratio
($\Phi_{\mbox{\scriptsize{2.0--8.0~keV}}}/\Phi_{\mbox{\scriptsize{0.5--2.0~keV}}}$)
is calculated to be $<$ 0.9 for \udrops \ corresponding to an effective photon
index lower limit of $\Gamma > 0.8$; a consistent, less tightly constrained
photon index is also derived for our bright \bdrops.  This lower limit is
consistent with our assumed $\Gamma$~=~2 photon index, which is expected for
galaxies with high star formation activity.  The derived average 2.0--8.0~keV
X-ray luminosities for these LBGs are $\approx$~1.5 and 1.4 $\times$ 10$^{41}$
\xlum \ for \udrops \ and bright \bdrops, respectively, a factor of
$\approx$~5--10 times higher than for typical starburst galaxies in the local
Universe \citep[e.g., M~82;][]{Griff2000}.  Furthermore, our LBGs are a factor
of $\approx$ 2 less X-ray luminous than those studied by B01 and N02.  

Assuming the X-ray luminosity functions of our individually undetected LBGs at
high redshifts have similar functional forms to those of normal/starburst
galaxies in the local Universe, we can estimate the expected median X-ray
luminosities for these LBGs.  This is important because, for a luminosity
function with significant skewness, the median will differ from the mean.  We
have investigated this using the \citet[hereafter DJF92]{David92} sample of 71
normal/starburst galaxies in the local Universe and find the mean to median
X-ray luminosity ratio
\Lx$^{\mbox{\scriptsize{mean}}}$/\Lx$^{\mbox{\scriptsize{median}}}$ $\approx$
5.  The mean 2.0--8.0~keV X-ray luminosity for these 71 galaxies is $\approx$ 6
$\times$ 10$^{40}$ \xlum, a factor of $\approx$ 2.5 lower than the mean X-ray
luminosities of our \udrops \ and optically bright subset of \bdrops.  If we
assume the \Lx$^{\mbox{\scriptsize{mean}}}$/\Lx$^{\mbox{\scriptsize{median}}}$
ratio for our LBGs is similar to that of the DJF92 sample, the corresponding
median 2.0--8.0 keV X-ray luminosity would be $\approx$~3~$\times$~10$^{40}$ \xlum.

The strong $U$-dropout signal allows investigation of the contribution of
specific subsets to the photon statistics (see $\S$~\ref{subsets} for details
and Table~\ref{tab:maglist} for results).  We find that the optically luminous
\udrops \ contribute most of the X-ray flux (S/N$_{\mbox{\scriptsize{bright}}}$
$\approx$ 1.5 $\times$ S/N$_{\mbox{\scriptsize{dim}}}$).  It is possible that
the emission from rest-frame UV may be heavily obscured by dust as would be
expected if LBGs were a ``scaled-up'' population of ultraluminous infrared
galaxies \citep[ULIRGs;][]{Golander2002}.  The X-ray to $B$-band mean
luminosity ratios for these LBGs suggests that this is likely not the case, and
the intrinsic dust attenuation from LBGs is similar to that expected for
star-forming galaxies \citep[see][]{Seibert2002}.  We find that relatively
asymmetric sources ($A$ $>$ \Amin) dominate the photon counts.  However, the
rest-frame $B$-band luminosities of these LBGs are somewhat elevated as
compared with sources of low-asymmetry index (i.e., $A$ $\le$ \Amin), and we
therefore offer no further interpretation.  Furthermore, stacking of the two
LBG subsets split by concentration index ($C$~$>$~\Cmin \ and $C$~$\le$~\Cmin)
show that these two morphological subsets have similar X-ray properties.  No
additional information was extracted from our divisions on asymmetry and
concentration.

Recent investigations have found that hard X-ray emission largely associated
with HMXBs can be used as a direct indicator of SFR
\citep[e.g.,][]{Bauer2002,Ranalli2003,Grimm2003,Persic2004}.  If we assume the
majority of the X-ray emission from our stacked detections of \udrops \ and
bright \bdrops \ is unobscured X-ray emisssion from HMXBs, we expect the
corresponding mean SFRs to be $\approx$~85--240 \sfr.  These SFRs were derived
using the linear SFR--$L_{\mbox{\scriptsize{2--10
keV}}}^{\mbox{\scriptsize{HMXB}}}$ relation given as equation (2) of
\citet{Persic2004} assuming a correlation error of 20$\%$.  For comparison, the
mean SFRs derived from the rest-frame 1500~\AA \ fluxes are estimated to be
$\approx$~65 and 10 \sfr \ for our \udrops \ and bright \bdrops, respectively,
without UV extinction.   When a correction is made for dust extinction, the
corresponding SFRs are $\approx$~400 and 60 \sfr \ (see G04).  The reasonable
agreement between the X-ray and UV extinction-corrected derived SFRs broadly
supports our assumption that the X-ray emission is dominated by star-forming
processes with little contribution needed from LLAGN. 

\begin{figure}
\centerline{
\includegraphics[width=9.5cm,angle=0]{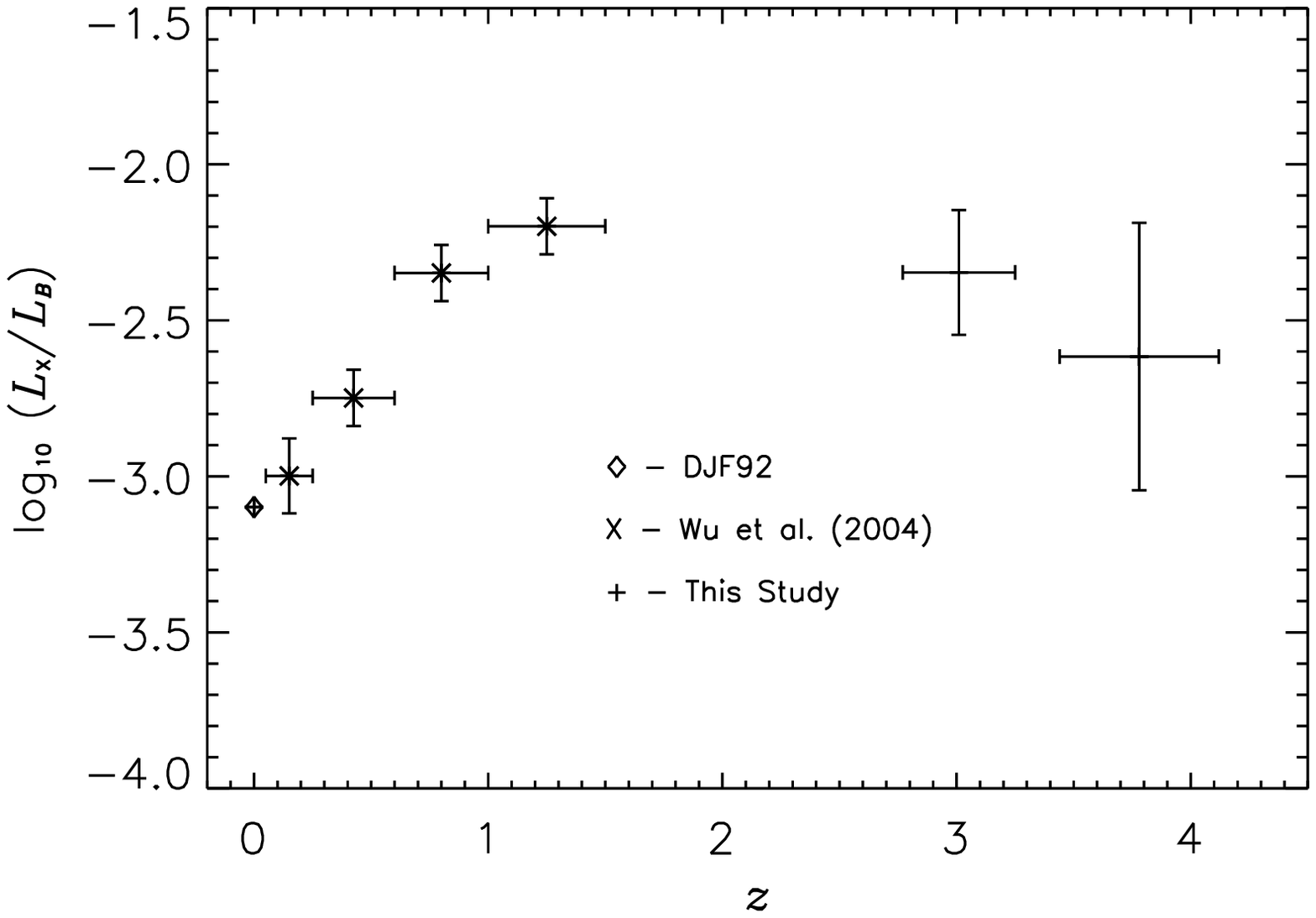}
\hfill
\includegraphics[width=9.5cm,angle=0]{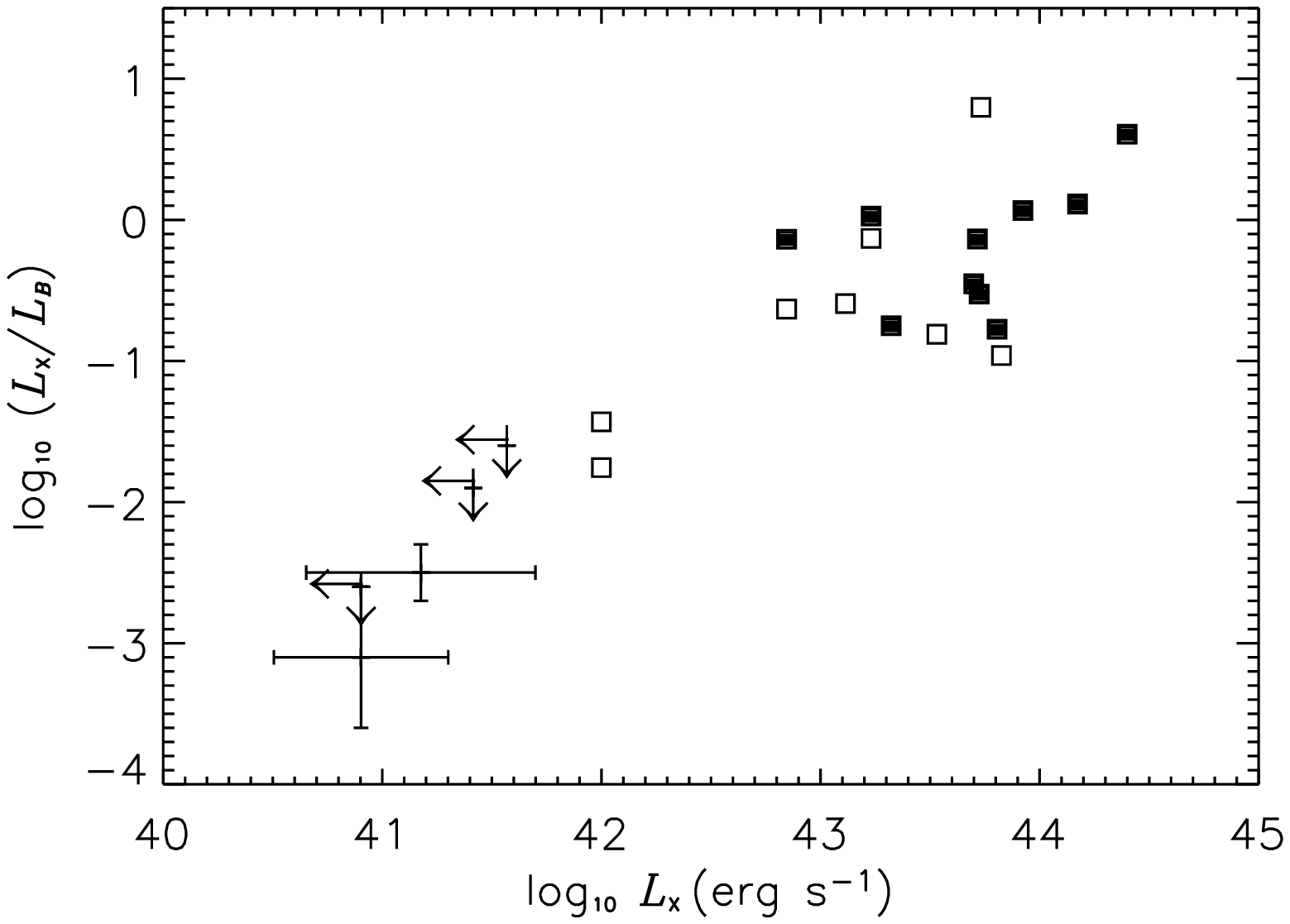}
}
\figcaption[fig5.eps]{({\it a}) The X-ray (0.5--4.5~keV) to $B$-band
luminosity ratio vs. redshift.  Note the suggested maximum of \ratio \ at $z$
$\sim$ 1.5--3 is consistent with predictions from the cosmic star formation
history.  Data used in this figure come from the following sources: $z$ $\sim$
0 (DJF92), $z$~$\sim$~0.05--1.5 \citep{Wu2004}, $z$ $\sim$ 3 and 4 -- our
\udrops \ and \bdrops, respectively. ({\it b}) X-ray to $B$-band luminosity
ratio vs. X-ray luminosities of individually undetected stacked LBGs (+) and
individually detected LBGs ($\sq$); solid squares represent broad-line AGN.
The individually detected sources are AGN (with one possible exception) and
have much larger \ratio \ ratios than our stacked LBGs.  X-ray luminosities
were calculated assuming a power law with photon index $\Gamma$ = 2, and
$B$-band luminosities were computed using $K$-corrections with an LBG SED (see
$\S$~\ref{subsets}).
\label{fig:dis}}
\end{figure}

The X-ray luminosities for star-forming galaxies (such as LBGs) are potential
tracers of global star formation history (Lilly~et~al. 1996; Madau~et~al.
1996; Blain~et~al. 1999).  A global estimate of the SFR can be implicitly drawn
from these LBGs by considering the mean X-ray luminosity per mean $B$-band
luminosity, \ratio \ (mean quantities).\footnote{In our stacking analyses we
compute an average $L_{\mbox{\tiny{X}}}$ for a given stack of source positions.
We therefore refer to $L_{\mbox{\tiny{X}}}/L_B$ as meaning
$<L_{\mbox{\tiny{X}}}>/<L_B>$, where $<L_{\mbox{\tiny{X}}}>$ and $<L_B>$ are
mean quantities.  When necessary for comparisons, data regarding
$L_{\mbox{\tiny{X}}}/L_B$ have been converted to this form.}  This ratio should
be interpreted with some caution due to the possible non-linear relationship
between \Lx \ and $L_B$ (i.e.  \Lx \ $\propto$ $L_B^{\alpha}$).  The power-law
index ($\alpha$) for this relation has been reported to range from $\approx$~1
\citep[DJF92;][]{Fabbiano1985} to as high as $\approx$~1.5 \citep{Shapley2001}
for galaxies in the local Universe.  We find that \ratio \ shows suggestive
evidence for evolution with redshift as illustrated in Figure~\ref{fig:dis}{\it
a}.  Here our \udrops \ and bright \bdrops \ are plotted and compared with the
results for normal galaxies at $z$ $\approx$ 0 (DJF92) and early-type spirals
from $z$ $\approx$~0.05--1.5 \citep{Wu2004}.  From these data we infer a peak
in \ratio \ at $z$ $\approx$~1.5--3.0, consistent with predictions of GW01.
These results, while based on larger samples of galaxies, are consistent with
the near constancy of $L_{\mbox{\scriptsize{X}}}/L_{\mbox{\scriptsize{UV}}}$ at
$z$ $\sim$ 1 and $z$ $\sim$ 3 reported by N02 for Balmer break galaxies and
LBGs, respectively.  In this description the global SFR is expected to peak at
$z$ $\approx$ 2.5--3.5.  During this epoch, X-ray emission is observed due to
contributions from HMXBs and supernovae.  Shortly after the UV luminous
population evolves, X-ray emission will continue in the form of LMXBs resulting
in an increase in \ratio.

When stacking LBGs at higher redshift, $z$ $\sim$ 5 and 6 (\vmag- and \idrops,
respectively), we do not obtain significant detections.  The AGN contribution
may be slightly more significant for these LBGs where the {\it Chandra}
exposures are only capable of detecting individual objects with
\hbox{2.0--8.0~keV} luminosities $\ga$~10$^{43}$ \xlum \ (implying many Seyfert
type AGN might not be individually detected at these redshifts).  In light of
these limits we therefore place constraints on the average AGN content of these
LBGs.  The derived rest-frame \hbox{2.0--8.0~keV} luminosity upper limits
(3$\sigma$) for the \vmag- and \idrops \ are 2.8 and 7.1 $\times$ 10$^{41}$
\xlum.  Such upper limits are characteristic of bright starburst or
low-luminosity Seyfert type galaxies and are the tightest constraints yet to be
placed on the X-ray properties of LBGs at $z$ $\ga$ 5.  For comparison, X-ray
analyses of 44 LBGs at $z$~$\sim$~5 (Bremer~et~al.~2003) and 54 LBGs at
$z$~$\sim$~6 (Moustakas \& Immler~2004) constrain their average
\hbox{2.0--8.0~keV} X-ray luminosities to be less than 3 and 7 $\times$
10$^{42}$~\xlum~(3$\sigma$), respectively.

Figure~\ref{fig:dis}{\it b} shows the overall \ratio \ vs \Lx \ results for
both individually detected and stacked LBGs.  Stacking analyses show that the
individually undetected LBGs at $z$ $\sim$ 3, 4, 5, and 6 have \ratio \ ratios
characteristic of local starburst galaxies.  Our individually detected LBGs
have much larger \ratio \ ratios, which are expected for X-ray luminous AGN.  

\acknowledgements

We gratefully acknowledge support from STScI grant HST-GO-09425.26-A and NSF
CAREER award AST-9983783 (BDL,WNB), the Royal Society (DMA), and NSF grant
03-07582 (DPS).   We thank Stefan Immler and Wentao Wu for useful discussions
and development of software.

\tabletypesize{\scriptsize}
\rotate
\begin{deluxetable}{ccccccccccc}
\tablecaption{Individually Detected LBGs
\label{tab:detected}}
\tablehead{
\colhead{{\it Chandra} Name} &
\colhead{Dropout} &
\colhead{Pos. Er.} &
\colhead{Offset} &
\colhead{} &
\colhead{Counts} &
\colhead{$f_{\mbox{\tiny{0.5--2 keV}}}$} &
\colhead{$f_{\mbox{\tiny{2--8 keV}}}$} &
\colhead{} &
\colhead{$L_{\mbox{\tiny{2--8 keV}}}$} &
\colhead{} \\
\colhead{J2000} &
\colhead{Bandpass} &
\colhead{(arcsec)} &
\colhead{(arcsec)} &
\colhead{$z$} &
\colhead{SB} &
\colhead{($\times$ 10$^{-16}$ \flux)} &
\colhead{($\times$ 10$^{-16}$ \flux)} &
\colhead{$\Gamma$} &
\colhead{($\times$ 10$^{43}$erg s$^{-1}$)} &
\colhead{\zmag} \\
\colhead{(1)} &
\colhead{(2)} &
\colhead{(3)} &
\colhead{(4)} &
\colhead{(5)} &
\colhead{(6)} &
\colhead{(7)} &
\colhead{(8)} &
\colhead{(9)} &
\colhead{(10)} &
\colhead{(11)} \\
}
\tablewidth{0pt}
\startdata
J033201.6$-$274327 & \umag & 0.4 & 0.15 & 2.726$^s$ & 450.7 & 27.7 & 56.2 & 1.5 & 14.9 &  23.8 \\
J033209.4$-$274807 & \umag & 0.3 & 0.10 & 2.81$^s$ & 195.6 & 11.0 & 59.7 & 0.8 & 6.4 &  22.4 \\
J033218.2$-$275241 & \umag & 0.6 & 0.10 & 2.801$^s$ & 55.9  & 3.6  &  6.3 & 1.6 & 2.1 &  24.1 \\
J033222.2$-$274937 & \bmag & 0.6 & 0.19 & --- & 10.1 & 0.6 & $<$ 3.8 & 1.4 & 0.7 & 27.5 \\
J033229.8$-$275106 & \bmag & 0.6 & 0.14 & 3.700$^s$ & 53.5 & 3.1 & 31.8 & 0.4 & 3.4 & 24.8 \\
J033238.8$-$275122 & \bmag & 0.6 & 0.25 & --- & 16.2 & 1.1 & 8.9 & 0.6 & 1.3 & 26.3 \\
J033239.1$-$274439 & \bmag & 0.6 & 0.24 & --- & 76.2 & 4.6 & 21.0 & 0.9 & 5.4 & 28.0 \\
J033239.7$-$274851 & \bmag & 0.3 & 0.14 & 3.064$^s$ & 134.6 & 7.5 & 7.1 & 0.4 & 5.3 & 24.8 \\
J033242.8$-$274702 & \bmag & 0.6 & 0.13 & 3.193$^s$ & 112.0 & 6.3 & 16.4 & 1.3 & 5.0 &  25.2 \\
J033243.2$-$274914 & \umag & 0.3 & 0.02 & 1.92$^s$ & 621.4 & 36.6 & 61.1 & 1.7 & 8.4 &  24.2 \\
J033244.3$-$275251 & \umag & 0.7 & 0.29 & 3.471$^s$ & 85.4  & 5.4  & 6.8  & 1.9 & 5.2 &  24.2 \\
J033250.2$-$275252 & \bmag & 0.4 & 0.33 & 3.6$^p$ & 404.2 & 24.0 & 75.3 & 1.2 & 25.1 & 26.1 \\
J123621.0+621412 & \umag & 0.6 & 0.16 & 1.74$^p$ & 131.9 & 3.7 & 6.0 & 1.7 & 0.7 &  25.4 \\
J123642.2+620612 & \bmag & 0.4 & 0.39 & 0.70$^p$ & 126.4 & 3.9 & 12.6 & 1.2 & 0.1 & 26.4 \\
J123647.9+621020 & \vmag & 0.6 & 0.26 & --- & 28.1 & 0.8 & 19.4 & $>$ 0.3 & 1.7 & 27.3 \\
J123648.0+620941 & \vmag & 0.6 & 0.23 & 5.186$^s$ & 96.7 & 2.7 & 4.9 & 1.6 & 6.7 & 23.8 \\
J123701.6+621146 & \imag & 0.6 & 0.43 & 1.52$^p$ & 13.9 & 0.4 & $<$ 1.2 & 1.4 & 0.1 & 27.8 \\
J123714.3+621208 & \umag & 0.6 & 0.14 & 3.146$^s$ & 74.0 & 2.2 & 5.2 & 1.4 & 1.7 & 25.5 \\
\enddata
\tablecomments{Col.(1): {\it Chandra} source name. Col. (2): Observed ACS
dropout bandpass. Col.(3): Positional error as reported in A03. Col.(4):
Angular offset between positions determined by {\it Chandra} and ACS. Col.(5):
Spectroscopic (``s'' superscript) or photometric (``p'' superscript) redshift.
Col.(6): Soft-band X-ray counts. Col.(7): Soft-band flux. Col.(8): Hard-band
flux. Col.(9): Effective photon index.  Here a value of 1.4 was assumed when photon statistics were too low to determine accurate values. Col.(10): Hard-band X-ray luminosity. Col.(11):
\zmag-band magnitude.}
\end{deluxetable}

\tabletypesize{\scriptsize}
\begin{deluxetable}{cccccccccccccc}
\tablecaption{Stacking Results For Filtered Dropout
Lists\label{tab:maglist}}
\tablehead{
\colhead{} &
\colhead{} &
\colhead{} &
\colhead{} &
\colhead{} &
\colhead{} &
\colhead{} &
\colhead{} &
\colhead{$\Phi$} &
\colhead{$f_{\mbox{\tiny{X}}}$} &
\colhead{$L_{\mbox{\tiny{X}}}$} &
\colhead{$L_B$} &
\colhead{} \\
\colhead{Data Type} &
\colhead{$N$} &
\colhead{$S$} &
\colhead{$B$} &
\colhead{$\sigma$} &
\colhead{ S/N } &
\colhead{ E(Ms) } &
\colhead{$N_{\mbox{\tiny {trials}}}$ > $S$} &
\colhead{(10$^{-7}$ counts s$^{-1}$)} &
\colhead{(10$^{-18}$ \flux)} &
\colhead{(10$^{41}$ ergs s$^{-1}$)} &
\colhead{(10$^{43}$ ergs s$^{-1}$)} &
\colhead{log($L_{\mbox{\tiny{X}}}/L_B$)} \\
\colhead{(1)} &
\colhead{(2)} &
\colhead{(3)} &
\colhead{(4)} &
\colhead{(5)} &
\colhead{(6)} &
\colhead{(7)} &
\colhead{(8)} &
\colhead{(9)} &
\colhead{(10)} &
\colhead{(11)} &
\colhead{(12)} &
\colhead{(13)} 
}
\tablewidth{0pt}
\startdata
 &  &  &  &  &  &  & & \bf{\udrops} \ ($z$ $\sim$ 3) &  &  & & & \\
\\
\tableline \\
     General &  449 &  1351.0 &  1115.3 &  33.4 &   7.1 &   663.9 &     0 & 3.6 $\pm$ 0.7 & 1.9 $\pm$ 0.4 & 1.5 $\pm$ 0.3 &  5.3 & $-$2.6 $\pm$ 0.2 \\
      Bright &  201 &   589.2 &   462.4 &  21.5 &   5.9 &   274.8 &     0 & 4.6 $\pm$ 1.2 & 2.4 $\pm$ 0.6 & 1.9 $\pm$ 0.5 & 10.3 & $-$2.7 $\pm$ 0.3 \\
         Dim &  248 &   759.7 &   651.9 &  25.5 &   4.2 &   389.2 &     0 & 2.8 $\pm$ 1.0 & 1.5 $\pm$ 0.5 & 1.2 $\pm$ 0.4 &  3.1 & $-$2.4 $\pm$ 0.3 \\
   $C>2.7$ &  223 &   672.0 &   559.4 &  23.7 &   4.8 &   332.7 &       0 & 3.4 $\pm$ 1.1 & 1.8 $\pm$ 0.6 & 1.4 $\pm$ 0.4 &  4.5 & $-$2.5 $\pm$ 0.3 \\
 $C\le2.7$ &  226 &   676.8 &   555.2 &  23.6 &   5.2 &   331.2 &       0 & 3.7 $\pm$ 1.1 & 1.9 $\pm$ 0.6 & 1.5 $\pm$ 0.4 &  6.1 & $-$2.6 $\pm$ 0.3 \\
     $A>0.1$ &  213 &   671.3 &   524.1 &  22.9 &   6.4 &   312.7 &     0 & 4.7 $\pm$ 1.1 & 2.5 $\pm$ 0.6 & 2.0 $\pm$ 0.5 &  6.7 & $-$2.5 $\pm$ 0.2 \\
   $A\le0.1$ &  236 &   677.5 &   591.0 &  24.3 &   3.6 &   351.2 &     0 & 2.5 $\pm$ 1.0 & 1.3 $\pm$ 0.5 & 1.0 $\pm$ 0.4 &  4.3 & $-$2.6 $\pm$ 0.4 \\
\\
\tableline \\
 &  &  &  &  &  &  & & \bf{\bdrops} \ ($z$ $\sim$ 4)  &  &  & &  & \\
\\
\tableline \\
     General & 1734 &  3749.5 &  3625.7 &  60.2 &   2.1 &  2102.2 &   197 & $\la$ 1.2 & $\la$ 0.6 & $\la$ 0.9 &  2.9 & $\la$ $-$2.4 \\
      Bright &  395 &   925.5 &   832.9 &  28.9 &   3.2 &   485.5 &     6 & 1.9 $\pm$ 0.9 & 1.0 $\pm$ 0.5 & 1.4 $\pm$ 0.6 &  9.2 & $-$2.8 $\pm$ 0.5 \\
         Dim & 1339 &  2823.8 &  2793.6 &  52.9 &   0.6 &  1616.8 &  2783 & $\la$ 1.4 & $\la$ 0.7 & $\la$ 1.0 &  2.1 & $\la$ $-$2.7 \\

\\
\tableline \\
 &  &  &  &  &  &  & & \bf{\vdrops} \ ($z$ $\sim$ 5)  &  &  &  & & \\
\\
\tableline \\
     General &  629 &  1382.0 &  1373.7 &  37.1 &   0.2 &   804.9 &  4067 & $\la$ 2.0 & $\la$ 1.0 & $\la$ 2.8 &  2.6 & $\la$ $-$2.8 \\
\\
\tableline \\
 &  &   &  &  &  &  & & \bf{\idrops} \ ($z$ $\sim$ 6)  &  &  &  & & \\
\\
\tableline \\
     General &  247 &   533.3 &   505.5 &  22.5 &   1.2 &   296.0 &  1065 & $\la$ 3.3 & $\la$ 1.7 & $\la$ 7.1 &  3.7 & $\la$ $-$1.8 \\
\enddata
\tablecomments{
The data apply to stacking analyses in the soft band.  Col.(1): Description of
the LBG sample stacked.  ``Bright'' and ``Dim'' subsets were created by
splitting the ``General'' lists at rest-frame $B$-band luminosities $\approx$
5.3 and 5.5~$\times$~10$^{43}$ \xlum \ ($M_B$ $\approx$ -20.6 and -20.5) for
\udrops \ and \bdrops, respectively.  Col.(2): Number of sources being stacked.
Col.(3): X-ray source counts obtained from stacking.  Col.(4): Mean background
counts obtained from Monte Carlo simulations.  Col.(5): Poisson error of the
background $B^{0.5}$.  Col.(6): Signal-to-noise ratio $(S-B)/B^{0.5}$.
Col.(7): Total, stacked effective exposure time.  Col.(8): Number of Monte
Carlo trials (out of 10,000) that produced a background estimate $>$~$S$.
Col.(9): X-ray count rate.  Col.(10): X-ray (0.5--2.0 keV) flux.  Col.(11):
2--8 keV rest-frame luminosity.  Col.(12): Rest-frame $B$-band luminosity.
Col.(12): Logarithm of X-ray to $B$-band luminosity ratio.  }
\end{deluxetable}

\end{document}